\documentclass{elsart}
\usepackage{epsfig}

\def\Li#1#2{{\mathrm{Li}}_{#1}\left(#2\right)}

\def\ba{\begin{eqnarray}}
\def\ea{\end{eqnarray}}
\def\dd{{\mathrm d}}
\def\la{\mathrel{\mathpalette\fun <}}

\def\fun#1#2{\lower3.6pt\vbox{\baselineskip0pt\lineskip.9pt
  \ialign{$\mathsurround=0pt#1\hfil##\hfil$\crcr#2\crcr\sim\crcr}}}
\def\order#1{{\mathcal O}\left(#1\right)}

\begin{document}
\begin{frontmatter}

\title{One--Loop Corrections to Radiative Muon Decay}

\author[BLTP]{A.B. Arbuzov\thanksref{UofA}},
\ead{ arbuzov@thsun1.jinr.ru}
\author[BLTP,SamaraUNC]{E.S. Scherbakova}

\thanks[UofA]{A part of this work was
performed in University of Alberta, Edmonton, Canada}

\address[BLTP]{Bogoliubov Laboratory of Theoretical Physics, \\
JINR,\ Dubna, \ 141980 \ \  Russia }
\address[SamaraUNC]{Samara State University, Samara, 443011 Russia \\
and University Centre JINR, Dubna, 141980 Russia}

\begin{abstract}
One--loop QED corrections to the differential width of radiative muon decay 
are considered. Results can be used to analyze high statistics data of modern
and future experiments.
\end{abstract}


\end{frontmatter}

\section{Introduction}

Since the discovery of muon in 1937, the studies of its properties 
were always very important for the progress of the elementary particle 
physics. Nowadays, such precision observations like the muon life time 
and the muon
anomalous magnetic moment are important for the checks of the Standard
Model and searches for {\em new physics.\/} Besides many others, the process
of radiative muon decay, 
\ba \label{proc}
\mu^+\ \longrightarrow\ e^+\ +\ \nu_e\ +\ \bar{\nu}_\mu\ +\ \gamma,
\ea
is investigated in the modern experiments.
In particular, the set of data from the PIBETA $(\pi\beta)$ 
experiment~\cite{Frlez:2003vg} at the Paul-Scherrer Institute contains 
a considerable amount of these decays. Accurate measurements of the process 
provide interesting information about the structure of weak interactions. 

In this paper we construct an advanced theoretical prediction for 
the differential distribution of process~(\ref{proc}). Our calculations 
of radiative corrections (RC) allow to reduce the 
theoretical uncertainty. That makes it possible to perform precision comparisons  
with the experimental data and potentially look for {\em new physics} or rule out
certain extensions of the Standard Model. 
 
In the limit of small energy loss (carried away by the neutrinos), radiative
corrections to the process were considered in Ref.~\cite{Arbuzov:1998kr}. 
In this limit the standard decay produces a background to the searches for
the neutrinoless decay $\mu \to e\gamma$.

In this paper we will consider the general kinematics assuming that the
energies of the final state electron and photon are above of a certain 
threshold and the angle between their momenta is not small (see Sect.~\ref{HColl}).
The tree--level distribution and the notation are introduced in the next Section.
Then we consider different RC contributions. 
In Conclusions we present some numerical results
and estimate the theoretical uncertainty in description of the radiative 
muon decay.

\section{The tree--level distribution}

Within the Fermi model of four--fermion interaction, the differential 
width of radiative muon decay 
was first considered in Refs.~\cite{Behrends:1956mb,Kinoshita:1959ru}.
Accurate formulae including the terms suppressed by the factor
$(m_e/m_\mu)^2$ were recently presented in Ref.~\cite{Kuno:2001jp}.
We checked that their results coincide with the relevant contribution,
which have appeared in calculations of exact one--loop radiative corrections
to the muon decay spectrum~\cite{Arbuzov:2001ui}.   
At the Born level the differential distribution of the electrons 
and photons of the process~(\ref{proc}) has the form
\ba \label{Born}
&& \frac{\dd^6\Gamma^{\mu^{\pm}\to e^{\pm}\nu\bar{\nu}\gamma}}
{\dd x\;\dd y\;\dd^2\Omega_e\;\dd^2\Omega_\gamma}
= \Gamma_0\frac{\alpha}{64\pi^3y}\beta\biggl[ F(x,y,d) 
\mp \beta \vec{P}_\mu\hat{p}_e G(x,y,d)
\mp \beta \vec{P}_\mu\hat{p}_\gamma H(x,y,d) \biggr], 
\nonumber \\  &&
\Gamma_0 = \frac{G_F^2 m_\mu^5}{192\pi^3}\, ,  
\qquad
d = 1 - \beta c, \qquad \beta = \sqrt{1-\frac{m_e^2}{E_e^2}}\, ,
\ea
where 
$G_F$ is the Fermi coupling constant; $m_e$ and $m_\mu$ are the electron and
muon masses, respectively;
$\vec{P}_\mu$ is the muon polarization vector; $x$ and $y$ are the electron
and photon energy fractions in the muon rest reference frame, $x=2E_e/m_\mu$
and $y=2E_\gamma/m_\mu$; 
by $\hat{p}_e$ and $\hat{p}_\gamma$ we denote the unit vectors
in the directions of motion of the electron and photon, 
$\hat{p}_e = \vec{p}_e/|\vec{p}_e|$
and $\hat{p}_\gamma = \vec{p}_\gamma/|\vec{p}_\gamma|$;
$c=\cos(\widehat{\vec{p}_e\vec{p}}_\gamma)$.
Functions $F(x,y,d)$, $G(x,y,d)$, and $H(x,y,d)$ 
can be found in Appendix of Ref.~\cite{Kuno:2001jp}.

In what follows we will concentrate on the case of unpolarized
muon decay, since it is the one measured in the PIBETA experiment.
In the unpolarized case only three variables are relevant and 
the tree--level distribution can be represented as
\ba \label{unpBrn}
\frac{\dd^3\Gamma^{\mathrm{Born}}_{\mathrm{unpol.}}}{\dd x\;\dd y\;\dd c}
= \Gamma_0 \frac{\alpha}{8\pi y}\beta F(x,y,d). 
\ea

Model independent parameterization of four--fermion interaction
(see Particle Data Group~\cite{Fetscher:eg}) 
leads to the appearance of two additional contributions.
One of them is proportional to the difference $(1-4\rho/3)$,
which describes the deviation of the Michel parameter $\rho$ from
its value in the Standard Model. And the other one
contains parameter $\bar{\eta}$, which is a positive semi-definite 
quantity (see Ref.~\cite{Fetscher:1993ki})
\ba \label{etabar}
\bar{\eta} &=& (|g_{RL}^V|^2+|g_{LR}^V|^2)
+ \frac{1}{8}(|g_{LR}^S + 2g_{LR}^T|^2 + |g_{RL}^S + 2g_{RL}^T|^2)
+ 2(|g_{LR}^T|^2+|g_{RL}^T|^2),
\ea
where $g^{S,V,T}_{RL,LR}$ are the right-left (RL) and left-right (LR) 
coupling constants, which parameterize
non--standard scalar (S), vector (V) and tensor (T) four--fermion interactions. 
In principle, one can look also for other exotic interactions, {\it e.g.}, for
the ones mediated by antisymmetric tensor fields~\cite{Chizhov:2004xw}. 
Extraction of $\bar\eta$ from the experimental data potentially 
  can put strict limits 
on physics beyond the Standard Model.

\section{Radiative corrections
\label{SecRC}}

New precision experiments call for an adequate level of accuracy in theoretical
predictions within the Standard Model. Effects of higher orders of the perturbation 
theory become important. Here we will consider 
the first order QED radiative corrections.
As usually, we separate them into three parts:  
{\it i)} emission of an additional soft photon;
{\it ii)} effect due to one--loop virtual photonic correction;
{\it iii)} emission of an additional hard photon.
Note that all the relevant pure week corrections (like loop insertions into the 
$W$-propagator) are included into the $G_F$ coupling 
constant~\cite{Marciano:1988vm,vanRitbergen:2000fi}, 
which is measured directly from the muon lifetime.
Effects of strong interactions in the process under consideration 
  are negligible for the moment.
They start to appear only at the order $\order{\alpha^2}$ through hadronic 
vacuum polarization. 

\subsection{Soft Photon Contribution}

We assume, that emission of an additional soft photon
of energy below certain threshold is not distinguished by
the experiment from the tree--level process~(\ref{proc}).
The energy of the soft photon, $\omega_{2},$
is limited by the parameter $\Delta$:
\ba
\omega_{2} \leq \Delta \frac{m_\mu}{2}\, ,\qquad \Delta \ll 1.
\ea

The corresponding correction can be factorized out
in front of the tree--level differential distribution:
\ba \label{softm}
\frac{\dd^3\Gamma^{\mathrm{Soft}}_{\mathrm{unpol.}}}{\dd x\;\dd y\;\dd c}
 &=& \delta^{\mathrm{Soft}}
\frac{\dd^3\Gamma^{\mathrm{Born}}_{\mathrm{unpol.}}}{\dd x\;\dd y\;\dd c}\, ,
\nonumber \\
\delta^{\mathrm{Soft}} &=&  - \frac{\alpha}{2\pi}\biggl\{
2\biggl( 2\ln\Delta
+ L + \ln\frac{m_e^2}{\lambda^2} \biggr)\biggl[ 1
- \frac{1}{2\beta}l_\beta \biggr] 
+ \frac{1}{2\beta}l^2_{\beta}
- \frac{1}{\beta}l_{\beta} 
\nonumber \\
&+& \frac{2}{\beta}\Li{2}{\frac{2\beta}{1+\beta}}
- 2 \biggr\}, \qquad
l_\beta = \ln\frac{1+\beta}{1-\beta}\; , 
\ea
where $\lambda$ is a fictitious photon mass; 
$L$ is the so--called large logarithm, 
$L=\ln(m_\mu^2/m_e^2) \approx 10.66\, ;$
the dilogarithm and the Riemann zeta--function are defined as usual:
\ba
\Li{2}{x} = - \int\limits_{0}^{1}\dd y\frac{\ln(1-xy)}{y}\, , \qquad
\zeta(n) = \sum _{k=1}^{\infty}\frac{1}{k^n}\, ,\quad 
\zeta(2) = \frac{\pi^2}{6}\, .
\ea
Quantity $\delta^{\mathrm{Soft}}$ coincides with the corresponding
factor, arising in the correction to the non--radiative muon decay
(see {\it e.g.} Ref.~\cite{Arbuzov:2001ui}). Expression~(\ref{softm})
takes into account the dependence on the electron mass exactly.
Omitting small terms proportional to $(m_e/m_\mu)^2$, we get
\ba
\delta^{\mathrm{Soft}} &=& 
\frac{\alpha}{2\pi}\biggl\{ \frac{1}{2}L^2
- (L-2+2\ln x)\biggl(1-2\ln\Delta - \ln\frac{m_e^2}{\lambda^2}\biggr)
- 2\ln^2x 
\nonumber \\
&+&  4\ln x - 2\zeta(2) \biggr\} + \order{\frac{m_e^2}{m_\mu^2}}.
\ea

\subsection{One--loop virtual correction}

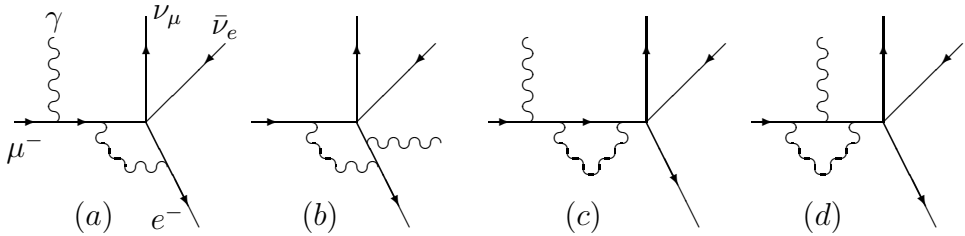
\begin{figure}[ht]
\unitlength=0.70mm
\special{em:linewidth 0.6pt}
\linethickness{0.6pt}
\begin{picture}(190.00,80.00)
\put(10.00,35.00){\line(1,0){25.00}}
\put(10.00,35.00){\vector(1,0){04.00}}
\put(14.00,35.00){\vector(1,0){10.00}}
\put(35.00,35.00){\line(0,1){20.00}}
\put(35.00,35.00){\vector(0,1){15.00}}
\put(35.00,35.00){\line(1,1){15.00}}
\put(50.00,50.00){\vector(-1,-1){04.00}}
\put(35.00,35.00){\line(1,-2){10.00}}
\put(35.00,35.00){\vector(1,-2){8.00}}
\put(17.50,36.00){\oval(2.00,2.00)[r]}
\put(17.50,38.00){\oval(2.00,2.00)[l]}
\put(17.50,40.00){\oval(2.00,2.00)[r]}
\put(17.50,42.00){\oval(2.00,2.00)[l]}
\put(17.50,44.00){\oval(2.00,2.00)[r]}
\put(17.50,46.00){\oval(2.00,2.00)[l]}
\put(17.50,48.00){\oval(2.00,2.00)[r]}
\put(17.50,50.00){\oval(2.00,2.00)[l]}
\put(26.50,34.00){\oval(2.00,2.00)[r]}
\put(26.50,32.00){\oval(2.00,2.00)[l]}
\put(26.50,29.50){\oval(3.00,3.00)[rt]}
\put(29.50,29.50){\oval(3.00,3.00)[lb]}
\put(29.50,26.50){\oval(3.00,3.00)[rt]}
\put(32.00,26.50){\oval(2.00,2.00)[b]}
\put(34.00,26.50){\oval(2.00,2.00)[t]}
\put(36.00,26.50){\oval(2.00,2.00)[b]}
\put(38.00,26.50){\oval(2.00,2.00)[t]}
\put(12.00,30.00){\makebox(0,0)[cc]{$\mu^-$}}
\put(17.50,54.00){\makebox(0,0)[cc]{$\gamma$}}
\put(39.00,55.00){\makebox(0,0)[cc]{$\nu_\mu$}}
\put(50.00,53.00){\makebox(0,0)[cc]{$\bar{\nu}_e$}}
\put(39.00,17.00){\makebox(0,0)[cc]{$e^-$}}
\put(25.00,17.00){\makebox(0,0)[cc]{$(a)$}}
\
\put(55.00,35.00){\line(1,0){20.00}}
\put(55.00,35.00){\vector(1,0){05.00}}
\put(75.00,35.00){\line(0,1){20.00}}
\put(75.00,35.00){\vector(0,1){15.00}}
\put(75.00,35.00){\line(1,1){15.00}}
\put(90.00,50.00){\vector(-1,-1){04.00}}
\put(75.00,35.00){\line(1,-2){10.00}}
\put(75.00,35.00){\vector(1,-2){8.00}}
\put(78.00,31.00){\oval(2.00,2.00)[t]}
\put(80.00,31.00){\oval(2.00,2.00)[b]}
\put(82.00,31.00){\oval(2.00,2.00)[t]}
\put(84.00,31.00){\oval(2.00,2.00)[b]}
\put(86.00,31.00){\oval(2.00,2.00)[t]}
\put(88.00,31.00){\oval(2.00,2.00)[b]}
\put(90.00,31.00){\oval(2.00,2.00)[t]}
\put(66.50,34.00){\oval(2.00,2.00)[r]}
\put(66.50,32.00){\oval(2.00,2.00)[l]}
\put(66.50,29.50){\oval(3.00,3.00)[rt]}
\put(69.50,29.50){\oval(3.00,3.00)[lb]}
\put(69.50,26.50){\oval(3.00,3.00)[rt]}
\put(72.00,26.50){\oval(2.00,2.00)[b]}
\put(74.00,26.50){\oval(2.00,2.00)[t]}
\put(76.00,26.50){\oval(2.00,2.00)[b]}
\put(78.00,26.50){\oval(2.00,2.00)[t]}
\put(68.00,17.00){\makebox(0,0)[cc]{$(b)$}}
\
\put(100.00,35.00){\line(1,0){30.00}}
\put(100.00,35.00){\vector(1,0){04.00}}
\put(104.00,35.00){\vector(1,0){16.00}}
\put(130.00,35.00){\line(0,1){20.00}}
\put(130.00,35.00){\vector(0,1){15.00}}
\put(130.00,35.00){\line(1,1){15.00}}
\put(145.00,50.00){\vector(-1,-1){04.00}}
\put(130.00,35.00){\line(1,-2){10.00}}
\put(130.00,35.00){\vector(1,-2){6.00}}
\put(107.50,36.00){\oval(2.00,2.00)[r]}
\put(107.50,38.00){\oval(2.00,2.00)[l]}
\put(107.50,40.00){\oval(2.00,2.00)[r]}
\put(107.50,42.00){\oval(2.00,2.00)[l]}
\put(107.50,44.00){\oval(2.00,2.00)[r]}
\put(107.50,46.00){\oval(2.00,2.00)[l]}
\put(107.50,48.00){\oval(2.00,2.00)[r]}
\put(107.50,50.00){\oval(2.00,2.00)[l]}

\put(113.50,34.00){\oval(2.00,2.00)[r]}
\put(113.50,32.00){\oval(2.00,2.00)[l]}
\put(113.50,29.50){\oval(3.00,3.00)[rt]}
\put(116.50,29.50){\oval(3.00,3.00)[lb]}
\put(116.50,26.50){\oval(3.00,3.00)[rt]}
\put(119.50,26.50){\oval(3.00,3.00)[b]}
\put(122.50,26.50){\oval(3.00,3.00)[lt]}
\put(122.50,29.50){\oval(3.00,3.00)[rb]}
\put(125.50,29.50){\oval(3.00,3.00)[lt]}
\put(125.50,32.00){\oval(2.00,2.00)[r]}
\put(125.50,34.00){\oval(2.00,2.00)[l]}
\put(118.00,17.00){\makebox(0,0)[cc]{$(c)$}}
\
\put(150.00,35.00){\line(1,0){25.00}}
\put(150.00,35.00){\vector(1,0){04.00}}
\put(175.00,35.00){\line(0,1){20.00}}
\put(175.00,35.00){\vector(0,1){15.00}}
\put(175.00,35.00){\line(1,1){15.00}}
\put(190.00,50.00){\vector(-1,-1){04.00}}
\put(175.00,35.00){\line(1,-2){10.00}}
\put(175.00,35.00){\vector(1,-2){8.00}}
\put(163.50,36.00){\oval(2.00,2.00)[r]}
\put(163.50,38.00){\oval(2.00,2.00)[l]}
\put(163.50,40.00){\oval(2.00,2.00)[r]}
\put(163.50,42.00){\oval(2.00,2.00)[l]}
\put(163.50,44.00){\oval(2.00,2.00)[r]}
\put(163.50,46.00){\oval(2.00,2.00)[l]}
\put(163.50,48.00){\oval(2.00,2.00)[r]}
\put(163.50,50.00){\oval(2.00,2.00)[l]}

\put(157.50,34.00){\oval(2.00,2.00)[r]}
\put(157.50,32.00){\oval(2.00,2.00)[l]}
\put(157.50,29.50){\oval(3.00,3.00)[rt]}
\put(160.50,29.50){\oval(3.00,3.00)[lb]}
\put(160.50,26.50){\oval(3.00,3.00)[rt]}
\put(163.50,26.50){\oval(3.00,3.00)[b]}
\put(166.50,26.50){\oval(3.00,3.00)[lt]}
\put(166.50,29.50){\oval(3.00,3.00)[rb]}
\put(169.50,29.50){\oval(3.00,3.00)[lt]}
\put(169.50,32.00){\oval(2.00,2.00)[r]}
\put(169.50,34.00){\oval(2.00,2.00)[l]}
\put(163.50,17.00){\makebox(0,0)[cc]{$(d)$}}
\end{picture}
\vspace{-.3cm}
\caption{Types of Feynman diagrams for radiative muon decay with one--loop RC.}
\label{Fig1}
\end{figure}

Here we will consider the effect of one--loop photonic
corrections. Some representatives of the relevant Feynman diagrams
are given in Fig.~\ref{Fig1}. There are two diagrams of class $(a)$ with photon
emission from an external leg (electron or muon line). In the same
way the two box-type diagrams of class $(b)$ describe real photon emission 
from virtual electron and muon propagators. Diagrams of classes $(c)$
and $(d)$ give corrections to photon radiation from a single leg.
To get the corresponding correction to the muon decay spectrum we
have to multiply the complete set of amplitudes of classes $(a-d)$
by two tree--level amplitudes, describing single photon emission.
In our calculations we followed the procedure which has been 
applied in Ref.~\cite{Kuraev:cg}.

\vspace{-.7cm}
\vspace{.3cm}

The standard technique for one--loop integration was used. 
The list of relevant integrals is given in Appendix~A.
To eliminate the 
ultraviolet divergences we applied renormalization of the masses and wave functions
of the electron and muon. Note that this is enough in the case of muon decay
(see Ref.~\cite{berman:1962xx,Sirlin:1980nh}), contrary to the general case of the 
Fermi four--fermion 
interaction. An analytical result for the virtual correction was obtained. 
We do not give the full formula here, since it is rather long.

\subsection{Emission of an additional collinear hard photon
\label{HColl}}

Events with registration of two hard photons are supposed to be rejected 
by the experimental event selection. But if the additional photon is
emitted at a small angle with respect to the momentum of
the outgoing electron (positron), the former is not recognized
by a calorimetric detector as an independent particle (this can
happen if there is no any considerable magnetic field in the 
detector volume). So, for the so--called collinear photon emission,
one observes an effective electron with the energy and momentum
composed by the sum of the corresponding quantities of the photon and
the {\em bare} electron. 
Let us assume that this kind of calorimetric registration happens in the experiment,
if the angle between the electron and photon momenta does not exceed
a certain value  $\theta_0$, which plays the role of a small parameter. 
We demand $m_e/m_\mu \ll \theta_0 \ll 1$. Typical experimental values for
this parameter, a few degrees, satisfy our conditions.
On the other hand, the angle between the observed photon and the electron 
should satisfy the condition $\theta=\widehat{\vec{p}_e\vec{p}}_\gamma \geq \theta_0$.

According to the general factorization procedure, we can represent the result 
for the contribution of collinear photon radiation as the product of two factors:
\ba \label{factor}
\frac{\dd^3\Gamma^{\mathrm{H-coll}}_{\mathrm{unpol.}}}{\dd x\;\dd y\;\dd c} &=&
\frac{\dd^3\Gamma^{\mathrm{Born}}_{\mathrm{unpol.}}}{\dd x\;\dd y\;\dd c}
R_{\mathrm{coll}}, 
\\ \nonumber  
R_{\mathrm{coll}} &=&
\frac{\alpha}{2\pi}\int\limits^{1}_{\Delta/x}
\frac{\dd z}{z}\, \biggl\{ [1+(1-z)^2]\biggl(L + 2\ln x - 1 
+ \ln\frac{\theta_0^2}{4} 
+ 2\ln(1-z)\biggr) + z^2 \biggr\}.
\ea
The tree--level radiative muon decay (with photon emission at large 
angles with respect to the electron momentum) serves
as a short--wave sub--process. 
Emission of a collinear photon by the outgoing electron
serves as a long--wave sub--processes. The formula for the collinear 
radiation factor agrees with the one in Ref.~\cite{Arbuzov:1997pj}.

Integration over the energy fraction of the collinear photon, $z$, gives
\ba \label{H-coll}
R_{\mathrm{coll}} &=& \frac{\alpha}{2\pi}\biggl[
\biggl(L + 2\ln x - 1 + \ln\frac{\theta_0^2}{4}\biggr)
\biggl(2\ln x - \frac{3}{2} - 2\ln\Delta\biggr)
- 4\zeta(2) + \frac{11}{4} \biggr].
\ea
Note that the lower limit of the collinear hard photon energy fraction 
is adjusted to the upper limit of soft photon emission.

\section{Results and conclusions}

Summing up the contributions of soft, virtual, and hard collinear photonic  
corrections we receive the final answer for the first order radiative 
correction to the process~(\ref{proc}).
Here is our result for the corrected distribution, which substitutes
function $F(x,y,d)$ from Eq.(\ref{unpBrn}):
\ba \label{Fcorr}
&& F^{\mathrm{Corr.}}(x,y,d) = F(x,y,d)\biggl(1+ \frac{\alpha}{2\pi}
A(x,y,d)\biggr) + \frac{\alpha}{2\pi}B_F(x,y,d),
\nonumber \\
&& A(x,y,d) = 2 \ln\frac{\theta_0^2}{4}\biggl( \ln x - \ln\Delta \biggr) 
           - 2 \ln\Delta 
           - \frac{3}{2} \ln\frac{\theta_0^2}{4}
           + \frac{1}{2}\biggl( \ln \frac{xyd}{2} - 2 \ln x \biggr)^2.
\ea 
We presented explicitly only the factorized part of the correction. 
The remaining non--factorizable part, $B_F(x,y,d)$, is rather long. 
We use it in a {\tt FORTRAN} code for numerical estimates. 
Expressions for the radiatively corrected functions $G^{\mathrm{Corr.}}(x,y,d)$
and $H^{\mathrm{Corr.}}(x,y,d)$ have exactly the same form as Eq.~(\ref{Fcorr})
with the trivial substitutions: $F\to G(H)$ and $B_F(x,y,d) \to B_{G(H)}(x,y,d)$.
The most important factorized part of the correction, $A(x,y,d)$, is
universal for all the three functions. 

It is worth to note that all the leading logarithm terms were factorized 
in each of the 
contributions, but they cancel out in the sum in accord with the 
Kinoshita--Lee--Nauenberg theorem~\cite{Kinoshita:1962ur,Lee:1964is}.
Moreover, all the dependence on the parameters $\Delta$ and $\theta_0$
is contained in $A(x,y,d)$.

\begin{figure}[ht]
\begin{center}
\includegraphics*[width=6cm,height=12cm,angle=270]{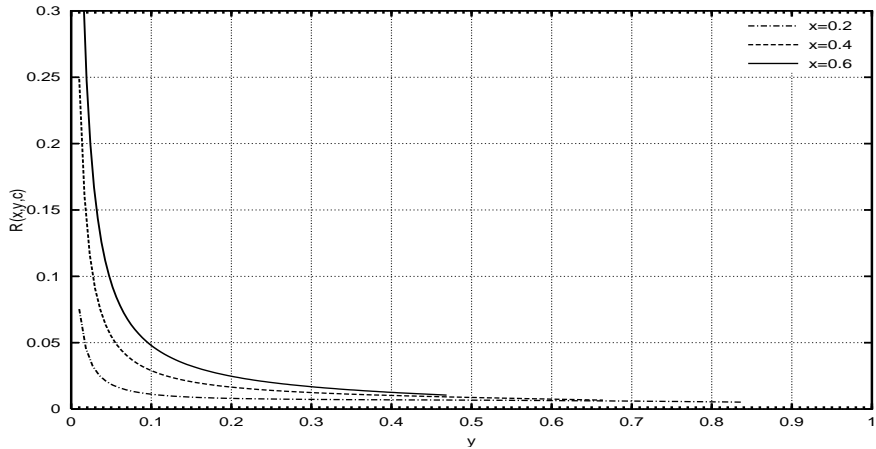}
\end{center}
\caption{Differential branching ratio {\it versus} electron energy fraction 
$y$ for three different $x$-values with fixed $c=0.5$, $\Delta=0.01$, 
$\theta_0=3^{\circ}$.} 
\label{Fig2}
\end{figure}

In Fig. \ref{Fig2} we plotted the Born--level differential branching ratio
of the radiative muon decay for a fixed value of $c$,
\ba
R(x,y,c) \equiv \frac{1}{\Gamma_0}\frac{\dd^3\Gamma^{\mathrm{Born}}_{\mathrm{unpol.}}}
{\dd x\;\dd y\;\dd c}\, .
\ea

\begin{figure}[ht]
\begin{center}
\includegraphics*[width=6cm,height=12cm,angle=270]{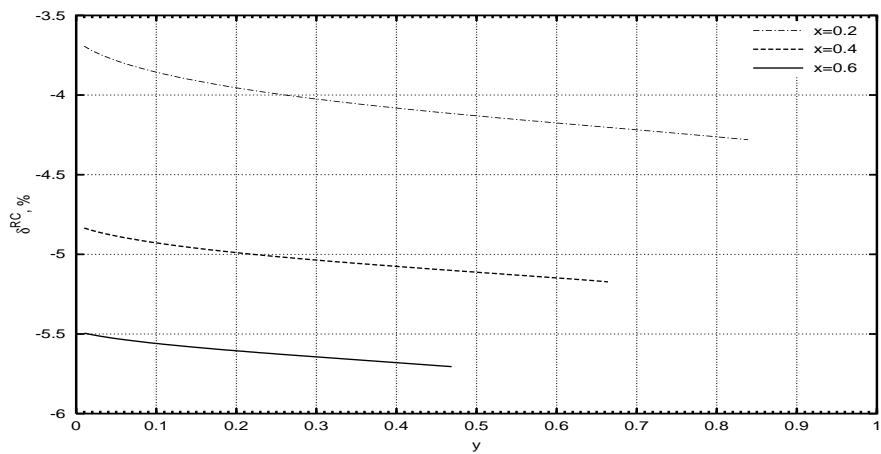}
\end{center}
\caption{Relative contribution of radiative corrections {\it versus}
the electron energy fraction; parameters are the same as in Fig.~\ref{Fig2}.}
\label{Fig3}
\end{figure}

The relative contribution of radiative corrections is illustrated by 
Fig.~\ref{Fig3},
\ba \label{deltrc}
\delta^{\mathrm{RC}} = \frac{F^{\mathrm{Corr.}}(x,y,d)-F(x,y,d)}
{F(x,y,d)}\cdot 100\%.
\ea
The dependence on $c$ value of $ \delta^{\mathrm{RC}}$ is rather weak.
For given values of $c$ and $x$ the maximal value of $y$ is defined
by the kinematics:
\ba
y_{\mathrm{max}} = \frac{1-x + m_e^2/m_\mu^2}{1-x(1-\beta c)/2}\, .
\ea
For the given set of parameters, the factorized part of the correction
dominates and gives about $4/5$ of the total effect.

\begin{figure}[ht]
\begin{center}
\includegraphics*[width=6cm,height=12cm,angle=270]{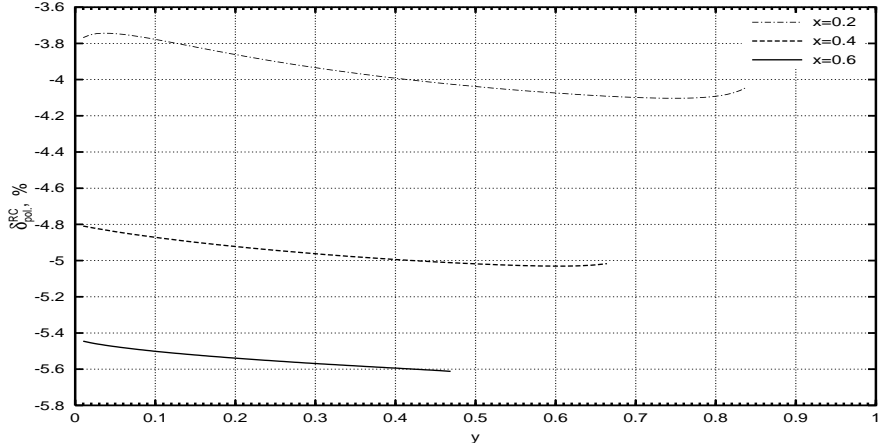}
\end{center}
\caption{Relative contribution of radiative corrections  for the case 
of polarized muon decay {\it versus} the electron energy fraction.}
\label{Fig4}
\end{figure}

To illustrate also the case of 100\% polarized muon decay 
we present in Fig.~\ref{Fig4} a plot for the relative contribution 
of radiative corrections for a set of fixed variables. 
Namely, $\widehat{\vec{P}_\mu \vec{p}}_e=30^\circ$,  
$\widehat{\vec{P}_\mu \vec{p}}_\gamma=60^\circ$;  
$c$, $\theta_0$ and $\Delta$ are the same as in Fig.~\ref{Fig2}.
Quantity $\delta^{\mathrm{RC}}_{\mathrm{pol.}}$ is defined 
in analogy to Eq.~(\ref{deltrc}) by adding the relevant contributions
of $G$ and $H$ functions according to Eq.~(\ref{Born}).

Thus we presented the calculation of one--loop QED corrections to the
differential distribution of unpolarized muon decay. Our {\tt FORTRAN}
code is available upon request from the authors.
The results can be applied also for the decays 
$\tau\to\mu\bar{\nu}_\mu\nu_\tau\gamma$ and 
$\tau\to e\bar{\nu}_e\nu_\tau\gamma$.
The theoretical uncertainty of the spectrum description is defined
by higher order QED radiative corrections (EW and QCD effects are
negligible compared to the QED ones). As a rough upper estimate we
can consider the relative contribution of the omitted higher 
order terms to be about $(\delta^{\mathrm{RC}})^2 \la 3\cdot 10^{-3}$,
which is small compared to the present experimental precision.
In principle, one can easily get the most important higher order terms
with logarithms of $\Delta$ and $\theta_0$ by means of soft and collinear
approximations.

\ack{
We thank V.~Baranov, E.~Kuraev, E.~Velicheva for fruitful discussions.
This work was supported by RFBR grant 04-02-17192.
One of us (E.Sch.) is grateful to the ``Dynasty'' foundation and ICFPM. 
}



\section*{Appendix A. \\ One--loop integrals}
\setcounter{equation}{0}
\renewcommand{\theequation}{A.\arabic{equation}}

Here we give the list of integrals over the loop momentum $k_1$,
which are used for calculations of the virtual loop contribution.
The notation for the loop integrals is as follows:
\ba
I^{1,\mu,\mu\nu}_{ijkl}
\equiv\int\frac{\dd^4k_1}{i\pi^2}\frac{1,k_1^{\mu},k_1^{\mu}k_1^{\nu}}
{(i)(j)(k)(l)}\, ,
\ea
where $(i-l)$ are the denominators of the relevant propagators:
\ba
&& (0) = k_1^2 - \lambda^2, \qquad
(1) = k_1^2 - 2k_1q_1, \qquad
(2) = k_1^2 - 2k_1p, \nonumber \\
&& (3) = k_1^2 - 2k_1(q_1+k_2) + \chi_e, \qquad
(4) = k_1^2 - 2k_1(p-k_2) - \chi_\mu, \nonumber \\
&& \chi_e = 2k_2q_1, \qquad \chi_\mu = 2k_2p.
\ea

Tensor and vector integrals are decomposed as
\ba
I^{\mu\nu}_{ijkl} &=& g^{\mu\nu}I^{g}_{ijkl}
+ p^{\mu}p^{\nu}I^{pp}_{ijkl}
+ q_1^{\mu}q_1^{\nu}I^{qq}_{ijkl}
+ k_2^{\mu}k_2^{\nu}I^{kk}_{ijkl}
+ (p^{\mu}q_1^{\nu}+p^{\nu}q_1^{\mu})I^{pq}_{ijkl} \nonumber \\
&+& (p^{\mu}k_2^{\nu}+p^{\nu}k_2^{\mu})I^{pk}_{ijkl}
+ (q_1^{\mu}k_2^{\nu}+q_1^{\nu}k_2^{\mu})I^{qk}_{ijkl}, \nonumber \\
I^{\mu}_{ijkl} &=& p^{\mu}I^{p}_{ijkl}
+ q_1^{\mu}I^{q}_{ijkl}
+ k_2^{\mu}I^{k}_{ijkl}.
\ea
We need different integrals with 4, 3, and 2 propagators.
In the integrals below we dropped the dimension and put $m_\mu=1$
(the dimension can be restored by multiplying by the required power of
$m_\mu$). In some cases we express tensor and vector integrals
through a combination of more simple ones. These relations can
be obtained by multiplying the tensor and vector integrals by certain
particle momenta and subsequent cancellation of denominators, where
possible. 

Ultraviolet and infrared divergences are regularized by introduction
of a cut--off, $\Lambda$, and an auxiliary photon mass, $\lambda$. They appear in the integrals
in two logarithms:
\ba
&& L_{\Lambda} = \ln\frac{\Lambda}{m_\mu^2}, \qquad \Lambda \gg m_\mu
\\
&& L_{\lambda} = \ln\frac{\lambda}{m_\mu^2}, \qquad \lambda \ll m_e.
\ea

The short notation $z=\frac{1}{2}\,x\,y\,d$ will be used below.
The relevant tensor integrals are
\ba
 && I_{013}^g = \frac{1}{4}\biggl( L_\Lambda + L - 1 - z I_{013}^{qk} \biggr),
\quad
 I_{013}^{qq} = - \frac{1}{z}\biggl( - z I_{013}^q + I_{01}^q - I_{03}^q \biggr),
\nonumber \\
 &&  I_{013}^{qk} = - \frac{1}{z}\biggl( L_\Lambda  + L - 1 - 2 z I_{013}^k 
- 2 I_{03}^q \biggr), 
\quad
I_{013}^{kk} = - \frac{1}{z}\biggl( I_{03}^q - I_{13}^k \biggr), 
\nonumber \\
 && I_{014}^g = \frac{1}{4}\biggl( I_{14} - (1-y) I_{014}^{pp} - (x-z) I_{014}^{pq} \biggr),
\nonumber \\
 && I_{014}^{qq} = \frac{1}{x-z}\biggl[
 + \frac{1}{2}  - \frac{L}{2} + \frac{y(L-1)}{x-z} ) - \frac{1}{2(1-x-y+z)} 
\nonumber \\ &&    \quad
 + \frac{ y^2}{(x-z)^2} \biggl( \Li{2}{\frac{y+z-z}{y}} 
 - \Li{2}{\frac{y-1}{y}} - \Li{2}{\frac{x+y-z-1}{x-z}}
\nonumber \\ &&    \quad
 + \Li{2}{\frac{z-x}{y}}
 - \Li{2}{\frac{1-y}{x-z}} 
 -  \Li{2}{1-x-y+z} 
\nonumber \\ &&    \quad
 - \frac{1}{2}\ln^2(x+y-z) - \ln(1-y)\ln(1-x-y+z) 
 + \ln(1-y)\ln(x - z) 
\nonumber \\ &&    \quad
 - L \ln(x+y-z)  + L \ln y - \ln^2(x+y-z)
  - \ln(x - z)\ln(1-x-y+z)  
\nonumber \\ &&    \quad
 - \ln(x - z)\ln(x+y-z)
 - \ln^2(x - z) + \ln y \ln(x+y-z))
\nonumber \\ &&    \quad
 + \ln y\ln(x - z)
 + \ln^2 y
 - 3 \ln y 
 + 2 \ln x
\biggr)
\nonumber \\ &&    \quad
 + \ln(x+y-z) \biggl( - \frac{1}{2} + \frac{y}{x-z} 
    + \frac{3y^2}{(x-z)^2} + \frac{y}{(1-x-y+z)(1-y)} 
\nonumber \\ &&    \quad
    + \frac{y}{(x-z)(1-y)}
    - \frac{1}{2(1-x-y+z)^2} \biggr)
 \biggr],
\nonumber \\
 && I_{014}^{pq} = \frac{1}{2(x-z)}\biggl[ \frac{1}{1-x-y+z} - 1
   + \ln(x+y-z)\biggl( \frac{y}{x-z} -  \frac{1}{(x-z)(1-y)} 
\nonumber \\ &&    \quad
+ \frac{1}{x-z} 
+  \frac{1}{(1-x-y+z)^2} - \frac{1}{(1-x-y+z)(1-y)} \biggr)
\nonumber \\ &&    \quad
+ \ln y \biggl(  - \frac{y}{x-z} + \frac{1}{(x-z)(1-y)}
- \frac{1}{x-z} \biggr)\biggr],
\nonumber \\
 && I_{014}^{pp} = \frac{1}{2(x-z)}\biggl[
 - \ln(x+y-z)\biggl( 1 - \frac{1}{1-x-y+z}\biggr)^2
+ \ln y \biggl( 1 - \frac{1}{1-y} \biggr)^2
\nonumber \\ &&  \quad
- \frac{1}{1-x-y+z} + \frac{1}{1-y} 
\biggr],
\nonumber \\
 && I_{023}^{g} = \frac{1}{4}\biggl( I_{23} - I_{023}^{pp} - z I_{023}^{qq} 
- (x+y) I_{023}^{pq} \biggr),
\nonumber \\
 && I_{023}^{qq} = - \frac{1}{x+y}\biggl( I_{03}^{q} - I_{23}^{q} 
        + 2 I_{023}^{pq} \biggr),  \quad
 I_{023}^{pp} = \frac{1}{x+y}\biggl( z I_{023}^{p} - 2 z I_{023}^{pq} 
- I_{02}^{p} +  I_{23}^{p} \biggr),
\nonumber \\
 && I_{023}^{pq} = -\frac{1}{(x+y)^2-4 z}\biggl( I_{23} (x+y) 
  - I_{23}^{p} ( 2 x + 2 y - 3 )
  + 3 z I_{023}^{p} + z I_{03}^{q} - z I_{23}^{q} 
- 3 I_{02}^{p} \biggr),
\nonumber \\
 && I_{024}^{g} = \frac{1}{4}\biggl( I_{24} - I_{024}^{pp} - y I_{024}^{pk} \biggr),
\quad
I_{024}^{kk} = \frac{1}{y}\biggl( I_{24}^{k} + I_{04}^{p} - 2 I_{024}^{pk} \biggr),
\nonumber \\
 && I_{024}^{pk} = \frac{1}{y}\biggl( 2I_{04}^{p} + 2 y I_{024}^{k} - I_{24} 
  + I_{024}^{pp} \biggr),
\quad
I_{024}^{pp} = \frac{1}{y}\biggl( y I_{024}^{p} + I_{02}^{p} - I_{04}^{p} \biggr),
\nonumber \\   
 && I_{123}^{g} = \frac{1}{4} \biggl[ \frac{3}{2} + L_{\Lambda} 
  + \frac{\ln x}{y-z} \biggl( x - \frac{1}{1-x}  + 1 \biggr) 
\nonumber \\ &&  \quad
 + \ln(x-z+y) \biggl( - 1 - \frac{x}{y-z} + \frac{1}{(y-z)(1-x-y+z)}
 - \frac{1}{y-z} \biggr) \biggr],
\nonumber \\
 && I_{123}^{pp} = \frac{1}{2(y-z)}\biggl[
 -  \ln(x+y-z)\biggl( 1 - \frac{1}{1-x-y+z}\biggr)^2
  +  \ln x\biggl( 1 - \frac{1}{1-x}\biggr)^2
\nonumber \\ && \quad
  - \frac{1}{1-x-y+z} + \frac{1}{1-x} \biggr],
\nonumber \\
 && I_{123}^{qq} = \frac{1}{2(y-z)}\biggl[ - 2 \ln(x+y-z)^2
  + \ln x\biggl( 2 L - 2 - \biggl( 1 - \frac{1}{1-x}\biggr)^2 \biggr)
\nonumber \\ && \quad
  + \ln(x+y-z)\biggl( 4 - 2 L - \biggl(1+\frac{1}{1-x-y+z}\biggr)^2\biggr)
\nonumber \\ && \quad
  + 2 \ln^2 x + 2 \Li{2}{1-x} - 2 \Li{2}{1-x-y+z} 
  - \frac{1}{1-x-y+z} + \frac{1}{1-x} \biggr],
\nonumber \\
 && I_{123}^{kk} = \frac{1}{y-z}\biggl[  1 - \frac{1}{2(1-x-y+z)}
 + \frac{x}{y-z}\biggl( L - 3 \biggr)
 - \frac{1}{2}L 
\nonumber \\ && \quad
  + \frac{x^2}{(y-z)^2}\biggl( ( L- 3) \biggl( \ln x - \ln(x+y-z) \biggr)
\nonumber \\ && \quad
 + \Li{2}{x} - \Li{2}{1-x-y+z}  - \ln^2(x+y-z) + \ln^2 x \biggr) 
\nonumber \\ && \quad
  + \frac{\ln(x+y-z)}{y-z}\biggl( - 1 + x + \frac{1}{1-x-y+z}\biggr)
\nonumber \\ && \quad
  - \frac{1}{2}\ln(x+y-z) \biggl( 1 + \frac{1}{1-x-y+z}\biggr)^2 \biggr],
\nonumber \\
 && I_{123}^{pq} = \frac{1}{2(y-z)}\biggl[
   \ln x\biggl( 1 - \frac{1}{(1-x)^2}\biggr) 
  - \ln(x+y-z)\biggl( 1 + \frac{1}{(1-x-y+z)^2}\biggr)
\nonumber \\ && \quad
  + \frac{1}{1-x-y+z} - \frac{1}{1-x} \biggr],
\nonumber \\
 && I_{123}^{pk} = \frac{1}{2(y-z)}\biggl[ \frac{1}{1-x-y+z}  - 1 
  + \frac{\ln x}{y-z}\biggl( \frac{1}{1-x} - 1 - x\biggr) 
\nonumber \\ && \quad
  + \frac{\ln(x+y-z)}{(1-x-y+z)^2}
  + \frac{\ln(x+y-z)}{y-z}\biggl( 1 + x - \frac{1}{1-x-y+z}\biggr)\biggr],
\nonumber \\
 && I_{123}^{qk} = \frac{1}{(y-z)}\biggl[\frac{5}{2} - L 
  - \frac{1}{2(1-x-y+z)} 
  + \frac{x}{y-z}\biggl( \frac{5}{2}\ln x - L\ln x - \ln x^2
\nonumber \\ && \quad
  + \ln(x+y-z)^2
   - \frac{1}{2(1-x)}\ln x - \Li{2}{1-x} 
      + \Li{2}{1-x-y+z} \biggr)\biggr],
\nonumber \\ && \quad
  + \frac{\ln(x+y-z)}{y-z}\biggl( Lx -\frac{1}{2} - \frac{5}{2}x 
                     + \frac{1}{2(1-x-y+z)}\biggr)
\nonumber \\ && \quad
  - \frac{1}{2} \ln(x+y-z)\biggl(1 + (1 + \frac{1}{1-x-y+z}\biggr)^2 \biggr],
\nonumber \\
 && I_{124}^{g} = \frac{1}{4(y-z)}\biggl[ (y-z)( L_{\Lambda} + \frac{3}{2}) 
  + 2 \Li{2}{1-x-y+z} - 2 \Li{2}{1-x}
\nonumber \\ && \quad
  + \ln x\biggl( x - 1 +\frac{1}{1-x}\biggr) 
  + \ln(x+y-z) \biggl( 1 - x - y + z - \frac{1}{1-x-y+z}\biggr)\biggr],
\nonumber \\
 && I_{124}^{pp} = \frac{1}{2(y-z)} \biggl[ \frac{1}{1-x-y+z} - \frac{1}{1-x}
 - \ln x \biggl( \biggl( 2 - \frac{1}{1-x}\biggr)^2 - 1\biggr)
\nonumber \\ && \quad
  + \ln(x+y-z) \biggl( \biggl( 2 - \frac{1}{1-x-y+z}\biggr)^2 - 1 \biggr)
   -2 \Li{2}{1-x}
\nonumber \\ && \quad 
 + 2 \Li{2}{1-x-y+z} \biggr],
\nonumber \\
 && I_{124}^{qq} = \frac{1}{2(y-z)} \biggl[ \frac{1}{1-x-y+z}
                          - \frac{1}{1-x}
  + \ln(x+y-z)\biggl( \frac{1}{(1-x-y+z)^2} - 1 \biggr)
\nonumber \\ && \quad
  + \ln x \biggl( 1 - \frac{1}{(1-x)^2} \biggr) \biggr],
\nonumber \\
 && I_{124}^{kk} = \frac{1}{(y-z)}\biggl[ 1+ \frac{1}{2(1-x-y+z)} 
  + \frac{6-3x}{y-z}
\nonumber \\ && \quad
  - \ln(x+y-z)\biggl( \frac{1}{2} + \frac{2}{1-x-y+z} 
                      - \frac{1}{2(1-x-y+z)^2}\biggr)
\nonumber \\ && \quad
  - \frac{\ln(x+y-z)}{y-z}\biggl( 5 - x  + \frac{1}{1-x-y+z}\biggr)
  + \frac{\ln(x+y-z)}{(y-z)^2}\biggl( 3x^2 - 6x \biggr)
\nonumber \\ && \quad
  + \frac{1}{(y-z)^2}\biggl( \ln x( 6 x - 3x^2 )  + (\Li{2}{1-x} 
   - \Li{2}{1-x-y+z})(6x - x^2 - 6)\biggr)\biggr],
\nonumber \\
 && I_{124}^{pq} = \frac{1}{2(y-z)}\biggl[ \frac{1}{(1-x)} 
                      - \frac{1}{(1-x-y+z)} 
 + \ln x\biggl( 1 - \frac{1}{1-x}\biggr)^2
\nonumber \\ && \quad
  - \ln(x+y-z)\biggl( 1 - \frac{1}{1-x-y+z}\biggr)^2\biggr],
\nonumber \\
 && I_{124}^{pk} = \frac{1}{2(y-z)}\biggl[ - 5 - \frac{1}{1-x-y+z} 
  + \ln(x+y-z)\biggl( 4 - \biggl( 2 - \frac{1}{1-x-y+z}\biggr)^2\biggr)
\nonumber \\ && \quad
  + \frac{\ln(x+y-z)}{y-z}\biggl( 5x - 1 + \frac{1}{(1-x-y+z)}\biggr)
  - \frac{\ln x}{y-z}\biggl( 5x - 1 + \frac{1}{1-x}\biggr) 
\nonumber \\ && \quad
  + \frac{6 - 2x}{y-z}\biggl( \Li{2}{1-x} - \Li{2}{1-x-y+z}\biggr)\biggr],
\nonumber \\
 && I_{124}^{qk} = \frac{1}{2(y-z)}\biggl[ 1 + \frac{1}{1-x-y+z} 
\nonumber \\ && \quad
  + \ln(x+y-z)\biggl( \frac{1}{(1-x-y+z)^2} - \frac{2}{(1-x-y+z)}\biggr)
\nonumber \\ && \quad
  + \frac{\ln(x+y-z)}{y-z}\biggl( 1 - x - \frac{1}{1-x-y+z}\biggr)
\nonumber \\ && \quad
  + \frac{1}{y-z}\biggl( \ln x( x - 1 + \frac{1}{1-x}) 
  - 2\Li{2}{1-x} 
+ 2\Li{2}{1-x-y+z}\biggr)\biggr]
\nonumber 
\ea

The following vector integrals were used in our calculations:
\ba
&& I_{0124}^k = \frac{1}{y}\biggl( -2I_{0124}^{p} - x I_{0124}^q 
                  + I_{124} - I_{014} \biggr),
\quad I_{0124}^q = \frac{1}{z}\biggl( -y I_{0124}^{p} + y I_{0124} 
                  + I_{012} - I_{014} \biggr),
\nonumber \\   
&& I_{0124}^{p} = \frac{1}{x^2 y^2} \biggl(
        - \frac{1}{2} \ln^2 x 
       + \ln(x - z + y)  \biggl( \ln(x-z) - \ln(1 - x + z - y)\biggr)
\nonumber \\  
&&    \quad   + \ln y \biggl(\ln x + \ln(1 - y) -  \ln(x - z) - \ln(x - z
+ y) \biggr)
       + \frac{1}{2} \ln^2 y - \zeta(2)
\nonumber \\  
&&    \quad  + \Li{2}{\frac{y+x-z}{y}} - \Li{2}{\frac{z-y}{x}}
       - \Li{2}{x - z + y} + \Li{2}{y}  - \Li{2}{1-x}
      \biggr),
\nonumber \\   
&& I_{0123}^k =  \frac{x}{zy}\biggl(  z I_{0123}^{q} - \frac{y}{x}\biggl(
I_{023} 
              - I_{123} \biggr) - z I_{0123} + I_{012} - I_{023} \biggr),
\nonumber \\   
&& I_{0123}^p = \frac{1}{x}\biggl( - z I_{0123}^{k} - I_{023} + I_{123}
\biggr),
\nonumber \\  
&& I_{0123}^{q} =  - \frac{1}{z( z - xy )}\biggl( \frac{1}{2} xyz I_{0123} 
              - \frac{1}{2} xy (I_{012} - I_{023}) - \frac{1}{2} yz (
I_{013} - I_{123})
\nonumber \\  
&&    \quad    + z (I_{012} - I_{023}) - z^2 I_{0123} + \frac{1}{2} y^2
(I_{023} - I_{123} )
             \biggr),
\nonumber \\ 
&& I_{012}^q = -\frac{1}{x}\biggl( L + \ln x + \frac{\ln x}{1-x}\biggr),
\quad
I_{012}^p = -\frac{1}{x}\biggl( \ln x - \frac{\ln x}{1-x}\biggr),
\nonumber \\
 && I_{013}^k = -\frac{1}{z}\biggl( z I_{013}^q - z I_{013} + I_{01} - I_{13}\biggr),
\quad  I_{013}^q = -\frac{1}{z}\biggl( - z I_{013} + I_{01} - I_{03}\biggr),
\nonumber \\
 && I_{014}^{q} = -\frac{y}{(x-z)^2}\biggl[ L \biggl( \frac{x-z}{y} 
            - \ln(x+y-z) + \ln y\biggr)
\nonumber \\ && \quad
  + \ln(x+y-z)\biggl( 2 - \frac{z-x+1}{y}
       + \frac{1}{y(1-x-y+z)}  - \frac{1}{1-x-y+z}\biggr) 
\nonumber \\ && \quad
- 2\ln y + 2\ln y\ln(x-z)
  - \ln y \ln(1-y) + 2\ln y \ln(x+y-z)  - \ln^2(x+y-z)
\nonumber \\ && \quad
   - 2\ln(x+y-z)\ln(x-z) - \ln^2(y)
   + \ln(x+y-z)\ln(1-x-y+z)
\nonumber \\ && \quad
  - 2\Li{2}{\frac{x+y-z}{y}} + \Li{2}{x+y-z} 
 - \Li{2}{y} + 2\zeta(2) \biggr],
\nonumber \\
 && I_{014}^{p} = - \frac{1}{x-z}\biggl[ \frac{y}{1-y}\ln y 
               - \frac{x+y-z}{1-x-y+z}\ln(x+y-z)\biggr],
\nonumber \\
 && I_{023}^p = -\frac{1}{2}\biggl( I_{023}^q (x+y) + I_{03} - I_{23}\biggr),
\nonumber \\
 && I_{023}^q = -\frac{1}{(x+y)^2-4z}\biggl( I_{03} (x+y) 
       - I_{23} (x+y-2) + 2 z I_{023} - 2 I_{02}\biggr),
\nonumber \\
 && I_{024}^{p} = - \frac{1}{y}\biggl[ \frac{y}{1-y}\ln y 
                 - \Li{2}{1-y} + \zeta(2) \biggr],
\nonumber \\
 && I_{024}^{k} = - \frac{1}{y}\biggl[ 2 - \ln y\biggl( 1 + \frac{1}{1-y}\biggr)
           + \frac{2}{y}\biggl( \Li{2}{1-y} - \zeta(2) \biggr)\biggr],
\nonumber \\
 && I_{123}^{p} = - \frac{1}{y-z}\biggl[ - \frac{x+y-z}{1-x-y+z}\ln(x+y-z)
           + \frac{x}{1-x}\ln x\biggr],
\nonumber \\
 && I_{123}^{q} = -\frac{1}{y-z}\biggl[ L\biggl( - \ln{x} + \ln(x+y-z)\biggr)
    + \ln^2(x+y-z)  -  \frac{x}{1-x}\ln x - \ln^2(x)
\nonumber \\ && \quad
   + \frac{x+y-z}{1-x-y+z} \ln(x+y-z) - \ln(x+y-z)\ln(1-x-y+z) 
\nonumber \\ && \quad
 - \Li{2}{1-x}
  - \Li{2}{x+y-z} + \zeta(2) \biggr],
\nonumber \\
 && I_{123}^{k} = \frac{1}{y-z}\biggl[ 2 + L \biggl( - 1 
                     - \frac{x}{y-z} \ln{x} 
                     + \frac{x}{y-z} \ln(x+y-z)\biggr)
\nonumber \\ && \quad
   + \frac{x}{y-z}\biggl(  \zeta(2)  
     - \ln(x+y-z)\ln(1-x-y+z)  + \ln^2(x-z+y)
\nonumber \\ && \quad
    - \Li{2}{x+y-z}  - \Li{2}{1-x}+ 2\ln x - \ln^2(x)\biggr)
\nonumber \\ && \quad 
  - \ln(x-z+y)\biggl( 1 + \frac{2x}{y-z} + \frac{1}{1-x-y+z}\biggr)
 \biggr],
\nonumber \\
 && I_{124}^{p} = - \frac{1}{y-z}\biggl[ - \frac{x}{1-x}\ln x
  + \frac{x+y-z}{1-x-y+z}\ln(x+y-z) + \Li{2}{1-x}
\nonumber \\ && \quad
  - \Li{2}{1-x-y+z}\biggr],
\nonumber \\
 && I_{124}^{q} = - \frac{1}{y-z}\biggl[ \frac{x}{1-x}\ln x
  - \frac{x+y-z}{1-x-y+z}\ln(x+y-z)\biggr],
\nonumber \\
 && I_{124}^{k} = - \frac{1}{y-z}\biggl[ 2 + \frac{x}{y-z}\ln x
  - \frac{x+y-z}{y-z}\ln(x+y-z)
\nonumber \\ && \quad
  + \frac{1-x}{y-z} \biggl(  \frac{x}{1-x}\ln x 
              - \frac{x+y-z}{1-x-y+z}\ln(x+y-z) \biggr)
\nonumber \\ && \quad
  + \frac{2-x}{y-z}\biggl( \Li{2}{1-x-y+z} - \Li{2}{1-x} \biggr)\biggr],
\nonumber \\
 && I_{01}^{q} = - \frac{1}{4} + \frac{1}{2}\biggl( L_{\Lambda} + L \biggr), 
 \quad
 I_{02}^{p} = - \frac{1}{4} + \frac{1}{2} L_{\Lambda},
\quad
I_{03}^{q} = \frac{1}{4} + \frac{1}{2}\biggl( L_{\Lambda} + L 
                                            - \ln z \biggr),
\nonumber \\
 && I_{04}^{p} = \frac{1}{4} + \frac{1}{2} \biggl( L_{\Lambda} 
             - \frac{y^2}{(1-y)^2} \ln y - \frac{1}{1-y} \biggr),
\nonumber \\
 && I_{12}^{q} = \frac{1}{4} + \frac{1}{2} \biggl( L_{\Lambda} 
             - \ln x + \frac{1}{(1-x)^2}\ln x + \frac{1}{1-x} \biggr),
\nonumber \\
 && I_{12}^{p} = \frac{1}{4} + \frac{1}{2} \biggl( L_{\Lambda} 
          - \frac{1}{1-x} - \frac{x^2}{(1-x)^2}\ln x\biggr),
\nonumber \\
 && I_{13}^q =   L_{\Lambda} + L - \frac{3}{2},
 \quad
I_{13}^k =  \frac{1}{2}\biggl( L_{\Lambda} + L - \frac{3}{2}\biggr),
 \quad
I_{14}^p = I_{14} - I_{14}^q - \frac{1}{2},  
\nonumber \\
 && I_{14}^q = I_{23}^q,  
 \quad
 I_{23}^p = I_{23} - I_{23}^q - \frac{1}{2},  
 \quad
I_{24}^p = L_{\Lambda} - \frac{3}{2},  
 \quad 
I_{24}^k = - \frac{1}{2}\biggl( L_{\Lambda} - \frac{3}{2}\biggr),
\nonumber \\
 && I_{23}^q = \frac{1}{4} + \frac{1}{2}\biggl[ L_{\Lambda} + \frac{1}{1-x-y+z}
  + \ln(x-z+y) \biggl( - 1 + \frac{1}{(1-x-y+z)^2} \biggr)\biggr],
\nonumber 
\ea
The scalar integrals read
\ba
 && I_{0123} = -\frac{1}{x z}\biggl[
  - \frac{1}{2}L_\lambda L - L_\lambda \ln x
  + \frac{1}{4}L^2
  - L \ln(x+y-z)
\nonumber \\ && \quad
  + L \ln x
  + L \ln z
  + 2\ln x \ln z
  - \ln^2(x-z+y)
  - 2\Li{2}{\frac{z-y}{x}}
  - \zeta(2)
\biggr],
\nonumber \\ 
 && I_{0124} = \frac{1}{x y}\biggl[
  - \frac{1}{2}L_\lambda L - L_\lambda \ln x
  - \frac{1}{4}L^2
  - L \ln(x-z+y)
\nonumber \\ && \quad
  + L \ln y
  + 2\ln x \ln y
  - \ln^2(x-z+y)
  - 2\Li{2}{\frac{z-y}{x}}
  - \zeta(2)
\biggr],
\nonumber \\ 
 && I_{012} = \frac{1}{x}\biggl[
\frac{1}{2}L L_\lambda + \ln x L_\lambda
+ \frac{1}{4}L^2 - \ln^2x - \Li{2}{1-x} \biggr],
\nonumber \\  && 
I_{013} = \frac{1}{z}\biggl[ \frac{1}{2}(L + \ln z)^2 - \zeta(2) \biggr],
\nonumber \\ 
 && I_{014} = - \frac{1}{x-z}\biggl[
L ( \ln(x+y-z) - \ln y )
  + \ln^2(x+y-z) + \ln y \ln(1 - y)
\nonumber \\ && \quad
  + \ln(x-z)\ln(x+y-z)
  -  \ln y \ln(x+y-z)
  - \ln y \ln(x-z)
\nonumber \\ && \quad
  + \Li{2}{\frac{x+y-z}{y}}
  + \Li{2}{1-x-y+z}
  - \Li{2}{ \frac{z-x}{y}}
  + \Li{2}{y}
  - 2\zeta(2)
\biggr],
\nonumber \\ 
 && I_{023} = \frac{1}{r(x+y)}\biggl[
      2\Li{2}{\frac{1+r}{2}(x+y) - \frac{1+r}{1-r}}
      - 2\Li{2}{1-\frac{2}{(1-r)(x+y)}}
\nonumber \\ && \quad
      - \Li{2}{1-x-y+z}
      + \frac{1}{2}\ln^2\frac{1+r}{2}
      - \frac{1}{2}\ln^2\frac{1-r}{2}
      - \ln\frac{1+r}{2} \ln\frac{1-r}{2}
\nonumber \\ && \quad
      - 2\ln(x+y) \ln\frac{1-r}{2}
      - \ln^2(x+y)
 \biggr],
\quad \quad r = \sqrt{1-\frac{4 z}{(x+y)^2}}\, , 
\nonumber \\ 
 && I_{024} = \frac{1}{y}\biggl[ \Li{2}{1-y} - \zeta(2) \biggr],
\nonumber \\ 
 && I_{123} = - \frac{1}{y-z}\biggl[
L (  \ln(x+y-z) - \ln x )
  + \zeta(2) 
 -  \Li{2}{x + y - z}
  - \Li{2}{1-x}
\nonumber \\ && \quad
  - \ln^2x
  - \ln(x+y-z)\ln(1-x-y+z)
  + \ln^2(x+y-z)
\biggr],
\nonumber \\
 && I_{124}  = - \frac{1}{y-z}\biggl[ \Li{2}{1-x} 
 - \Li{2}{1-x-y+z}  \biggr],
\nonumber \\
 && I_{01}  = L_{\Lambda} + L + 1, \quad
  I_{02}  = L_{\Lambda} +  1,\quad
  I_{03}  = L_{\Lambda} + L + 1 - \ln x,
\nonumber \\
 && I_{04}  = L_{\Lambda} + 1 + \frac{y}{1-y}\ln y,\quad
  I_{12}  = L_{\Lambda} + 1 + \frac{x}{1-x}\ln x,\quad
  I_{13}  = L_{\Lambda} + L - 1,
\nonumber \\
 && I_{14} = I_{23} = L_{\Lambda} + 1 + \frac{x+y-z}{1-x-y+z}\ln(x+y-z), 
\quad  I_{24} = L_{\Lambda} - 1. 
\nonumber
\ea

\end{document}